\DocumentMetadata{}   
\documentclass[sigconf]{acmart}

\copyrightyear{2025}
\acmYear{2025}
\setcopyright{cc}
\setcctype{by}
\acmConference[WiSec 2025]{18th ACM Conference on Security and Privacy in Wireless and Mobile Networks}{June 30-July 3, 2025}{Arlington, VA, USA}
\acmBooktitle{18th ACM Conference on Security and Privacy in Wireless and Mobile Networks (WiSec 2025), June 30-July 3, 2025, Arlington, VA, USA}
\acmDOI{10.1145/3734477.3734704}
\acmISBN{979-8-4007-1530-3/2025/06}

\usepackage{xspace}
\usepackage{xcolor}
\usepackage{multirow}
\usepackage{tabularx} 
\usepackage[tableposition=top]{caption}  
\usepackage[normalem]{ulem}
\usepackage{hhline}
\usepackage[absolute,overlay]{textpos}

\newcommand{\sysname}{\texttt{ARMOUR}\xspace}

\definecolor{revcol}{rgb}{0 0 0}
\newcommand{\rev}[1]{\textcolor{revcol}{#1}}

\newcommand*{\mline}[1]{%
\begingroup
    \renewcommand*{\arraystretch}{1.1}%
   \begin{tabular}[c]{@{}>{\raggedright\arraybackslash}p{9cm}@{}}#1\end{tabular}%
  \endgroup
}

\begin{document}

\title[ARMOUR US: Monitoring Android Zero-permission Sensor Usage From User Space]{ARMOUR US:
\uline{A}ndroid \uline{R}untime Zero-per\uline{m}ission Sens\uline{o}r \uline{U}sage Monito\uline{r}ing from \uline{U}ser \uline{S}pace}

\author{Yan Long}
\authornote{Co-first authors with equal contributions.}
\affiliation{%
  \institution{Northeastern University}
  \city{Boston}
  \state{Massachusetts}
  \country{USA}
}
\email{y.long@northeastern.edu}

\author{Jiancong Cui}
\authornotemark[1] 
\affiliation{%
  \institution{Northeastern University}
  \city{Boston}
  \state{Massachusetts}
  \country{USA}
}
\email{cui.jianc@northeastern.edu}

\author{Yuqing Yang}
\affiliation{%
  \institution{The Ohio State University}
  \city{Columbus}
  \state{Ohio}
  \country{USA}
}
\email{yang.5656@osu.edu}

\author{Tobias Alam}
\affiliation{%
  \institution{University of Michigan}
  \city{Ann Arbor}
  \state{Michigan}
  \country{USA}
}
\email{tobiasal@umich.edu}

\author{Zhiqiang Lin}
\affiliation{%
  \institution{The Ohio State University}
  \city{Columbus}
  \state{Ohio}
  \country{USA}
}
\email{zlin@cse.ohio-state.edu}

\author{Kevin Fu}
\affiliation{%
  \institution{Northeastern University}
  \city{Boston}
  \state{Massachusetts}
  \country{USA}
}
\email{k.fu@northeastern.edu}

\begin{abstract}
This work investigates how to monitor access to Android zero-permission sensors which could cause privacy leakage to users. Moreover, monitoring such sensitive access allows security researchers to characterize potential sensor abuse patterns. Zero-permission sensors such as accelerometers have become an indispensable part of Android devices. The critical information they provide has attracted extensive research investigating how data collectors could capture more sensor data to enable both benign and exploitative applications. In contrast, little work has explored how to enable data providers, such as end users, to understand sensor usage. While existing methods such as static analysis and hooking-based dynamic analysis face challenges of requiring complicated development chains, rooting privilege, and app-specific reverse engineering analysis, our work aims to bridge this gap by developing \sysname for user-space runtime monitoring, leveraging the intrinsic sampling rate variation and convergence behaviors of Android. \sysname enables privacy-aware users to easily monitor how third-party apps use sensor data and support security researchers to perform rapid app-agnostic sensor access analysis. 
Our evaluation with 1,448 commercial applications shows the effectiveness of \sysname in detecting sensor usage in obfuscated code and other conditions, and observes salient sensor abuse patterns such as 50\% of apps from seemingly sensor-independent categories accessing data of multiple zero-permission sensors. 
We analyze the impact of Android's recent policy changes on zero-permission sensors and remaining technical and regulatory problems.


\end{abstract}

\ccsdesc[500]{Security and privacy~Mobile and wireless security}
\ccsdesc[500]{Security and privacy~Malware and its mitigation}

\keywords{Android, zero-permission, sensors, runtime monitoring, privacy}

\maketitle

\section{Introduction}

\begin{figure}[!t]
\vspace{.1in}
	\centering
\includegraphics[width=.48\textwidth]{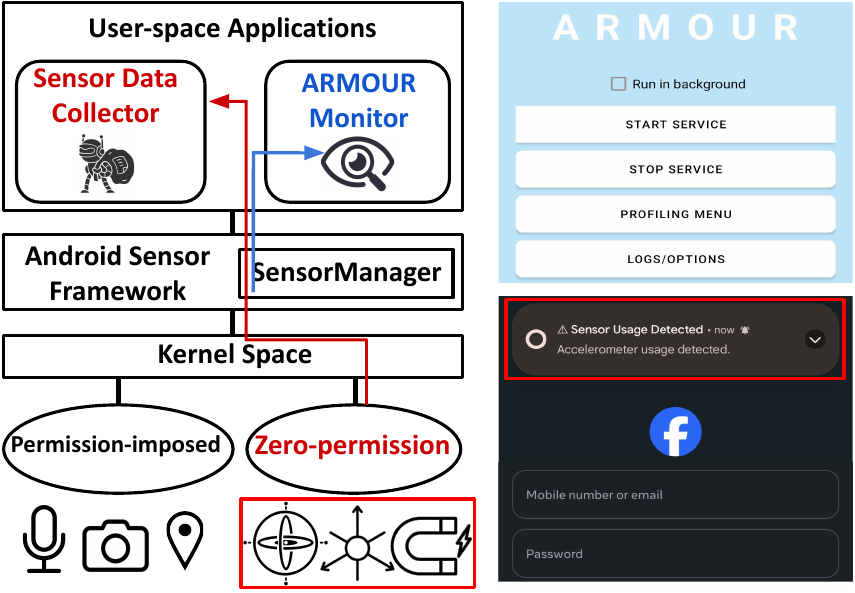}
\vspace{-.2in}
	\caption{Although third-party apps can collect zero-permission sensor data almost without regulations in existing Android ecosystems, \sysname provides a complementary defender capability of monitoring sensor access information solely from the user space.} 
	\label{fig:sensor_framework}
    \vspace{-.1in}
\end{figure}

Sensors in smartphones and wearable devices have become a cornerstone of the seamless interactions between the physical world and cyberspace. As of 2025, there are over 6.5 billion smartphone users worldwide with Android-powered devices taking up over 80\% of the market share~\cite{market}. Android devices are providing an increasing number of sensor hardware with growing capabilities to support software in acquiring more information from the physical world, enabling a wide range of applications such as health monitoring and activity recognition~\cite{harari2017smartphone, majumder2019smartphone, pennekamp2017survey, long2023side}. 
While Android allows third-party apps to access and analyze sensor data for diverse purposes, the current Android platform provides very limited mechanisms for monitoring and revealing third-party applications' sensor usage. In particular, applications can access readings of accelerometers, gyroscopes, magnetometers, and other types of sensors which can capture important motion and environmental information of the physical devices and users, without requiring user permissions or even notifying users of their usage. Such sensors are thus referred to as \textit{zero-permission sensors} (Figure~\ref{fig:sensor_framework}).

The unmonitored usage of zero-permission sensors creates nearly unregulated information channels that may leak sensitive information. This risk has manifested in a significant number of recent research works that demonstrate how zero-permission sensor data could contain critical private information. For example, smartphones' accelerometer and gyroscope readings contain vibration signals caused by human body movements and even sound waves of voice, which further enable the inference of users' age, identity, speech, location, and password inputs~\cite{michalevsky2014gyrophone, anand2019spearphone, ba2020learning, davarci2017age, lee2017implicit, manifest, spreitzer2014pin, owusu2012accessory, cai2011touchlogger, shamsi25EyeHearYou}. Another example is the feasibility of identifying various user-device interactions by analyzing the electromagnetic signals embedded in magnetometer readings that are generated by smartphone CPUs or displays~\cite{cheng2019magattack, pan2021magthief, mohamed2023istelan,long2024eye}. Research has also shown similar problems on emerging AR/VR devices, which could present more pervasive threats given the predicted continuous wearing of these devices~\cite{wu2023privacy, shi2021face}. \rev{While the possible threats identified by prior research have not yet been analyzed on commercial apps,} 
the mere fact that third-party apps have unmonitored access to information-rich sensor data has already raised significant concerns.

As a result, we observe an information asymmetry gap between data collectors (e.g., application developers) and data providers (e.g., application users): \textit{Despite extensive research investigating how to collect more sensor data to enable various applications, little work has explored how to meet the needs of researchers and data providers to monitor and analyze the sensor usage.} Although there exist commercial privacy-preserving applications \cite{safedot, privacydot} that support users in tracking the usage of permission-imposed cameras and microphones, none of them is capable of monitoring zero-permission sensors. This overlooked fundamental capability calls for the development of mechanisms that allow end users and researchers to understand the pattern of zero-permission sensor access. Several existing techniques may facilitate such analysis, but still suffer from limitations. On one hand, \textit{Static Analysis} methods may be used to detect codes accessing sensors but face challenges in three aspects: (1) lack of support for user-side runtime monitoring, (2) challenges in creating time-variant information channel,
and (3) lack of semantic information among obfuscated and native codes.
On the other hand, despite that \textit{Hooking-based Dynamic Analysis} methods may address these challenges to some degree, they still face three limitations:
(1) excessive expertise is required to use sophisticated software development chains, (2) physical devices with escalated privileges such as rooting are needed to perform monitoring, and (3) prior knowledge of and customization for each application needs to be obtained prior to analyzing them. 

Our work addresses these constraints and contributes complementary capabilities by developing a methodology and tool for \textit{accessible user-space runtime monitoring} of Android zero-permission sensor usage. Our tool \sysname\footnote{Code and dataset available at \href{https://github.com/longyan97/ARMOUR}{https://github.com/longyan97/ARMOUR}} is capable of detecting the time of access, type of sensor, and the sampling rates accessed by third-party applications. \sysname is a privilege-free, off-the-shelf Android module that can run as an individual background monitoring app (Figure~\ref{fig:sensor_framework}) or be seamlessly incorporated into other trusted privacy-preserving applications. It enables non-expert end users to easily monitor how third-party applications may be harvesting their sensor data and helps security researchers rapidly identify and characterize abuse patterns of zero-permission sensors. \sysname builds upon a key observation of the Android sensor framework's rule of sampling rate variation and convergence. Specifically, we observe that the actual instant sampling rates Android OS provides to different applications are interdependent, creating an information channel that can be leveraged by a trusted defender application to monitor the sensor usage of concurrent applications. \sysname running in the user space is thus able to detect sensor usage by other third-party applications by listening for sampling rate changes in its own sensor data packets. Our tests show that \sysname can detect sensor usage across diverse hardware and software platforms and is robust against native codes and obfuscations. Besides user-installed applications, it can also detect the usage of zero-permission sensors by websites. \rev{The utilization of this previously unexplored behavior of the Android sensor framework provides unique light-weight, user-space monitoring capabilities that enable users and security analysts to bridge the information asymmetry gap revealed by prior works.}

Our evaluation aims to both measure the performance of \sysname in terms of its detection accuracy and runtime overhead, and use \sysname to characterize real-world sensor usage of commercial applications. In the former, we collected 50 apps known to use sensors and found that \sysname could detect all the usage. 
Additional prolonged testing of no sensor activities confirmed \sysname's ability to avoid false positive reports.  When running continuously as a background monitoring application, \sysname only incurs an overhead of about 2.6\% phone battery usage per hour. In the latter, we built another dataset with 1,398 of the most popular Google Play applications in 35 categories and performed in-depth analyses, resulting in multiple insights. For example, our results show that certain app categories such as Books, Finance, and Shopping have unexpectedly high sensor usage without clear motivation and justification. Sensor-based tracking and identification services commonly used by Finance applications confirm that sensor data is capable of capturing privacy-sensitive information. By comparing sensor sampling rate distributions in older and newer Android versions, we verified the positive impact of the \texttt{HIGH\_SAMPLING\_RATE\_SENSORS} permission introduced in Android 12 in terms of security, but observed outstanding problems. For instance, apps may actively request the highest possible sampling rates. Moreover, device manufacturers may bypass Android's high sampling rate regulations, thus undermining the positive impact of the introduced permission. We also analyze case studies of abnormal sensor usage patterns measured by \sysname. For example, 3.6\% of the apps continue to access sensors on certain smartphones even when their GUI is terminated by users. 
\rev{Manual testing demonstrates additional sensor usage that can be triggered by GUI interactions such as user login, indicating that our work's result provides a lower-bound measurement that future works can build upon.} Overall, our measurements reveal obvious zero-permission sensor abuse problems that could motivate future research in providing better user privacy protections, refining sensor data access policies, and performing in-depth sensor usage attribution analysis. Our major contributions are summarized as:
\begin{itemize}
    \item The problem formulation of zero-permission sensor usage monitoring from the defender standpoint and the first proposed mechanism for user-space runtime detection. The tool \sysname developed for end users and researchers complements existing Android OS and static/dynamic analysis capabilities, making a key step toward accessible sensor data privacy controls.  
    \item A dedicated Android app dataset collected for zero-permission sensor usage analysis that consists of 1,448 popular apps. The open-source dataset's results provide ground-truth and baseline measurements that other works of Android sensor security and privacy can compare with. 
    \item The detailed analysis of existing sensor abuse patterns in popular commercial Android applications. Our observations highlight abnormal and unjustified sensor data access, opening up possible research venues for improving the technical control and policy regulations of zero-permission sensor data access. 
\end{itemize}

\section{Background}

This section provides the necessary background for understanding the current status and remaining gaps of zero-permission problems.



\subsection{Android Sensor Framework}\label{sec:sensor_framework}
Figure~\ref{fig:sensor_framework} provides an overview of the Android sensor framework. Android provides two main categories of sensors, namely permission-imposed sensors including cameras, microphones, GPS, etc., and zero-permission sensors. When third-party applications want to access permission-imposed sensors, users will be prompted to decide whether to allow the sensor usage. Zero-permission sensors, on the other hand, can be used without requiring installation-time or run-time consent.

\textbf{Zero-permission Sensors.} 
Table~\ref{tab:sensor_table} lists the main categories and representative instances of zero-permission sensors. Motion sensors such as accelerometers and gyroscopes can collect information about the movement of the device, which has been shown to reflect the physical activities of users such as walking, typing, etc. Position sensors such as magnetometers collect information about the device's physical positions and can also embed the electromagnetic characteristics of the smartphone's surroundings.

\textbf{Risks and Existing Countermeasures.} Although previously deemed non-sensitive, research in the past few years has proven that such zero-permission sensors' data actually contain significantly more critical and fine-grained information than what users and smartphone OS designers expected and can be exploited by third-party applications to identify smartphone users, steal touchscreen inputs, etc. Section~\ref{sec:related_exploitation} provides more details about the privacy-sensitive information zero-permission sensor data could contain.

Android made a series of adjustments to its policies to counteract these emerging threats discovered by previous research. Starting Android 9, apps are required to run in foreground services to access sensor data in the background~\cite{android9}. Foreground services create visible dialogues on the menu bar to notify users that the application is running in the background. Nevertheless, there is still no way of knowing if apps are collecting sensor data. Android 12 starts to limit the highest available sampling rate to 200 Hz for common sensor usage by third-party apps, where any app that needs a higher sampling rate is required to declare the use in the manifest file through the \texttt{HIGH\_SAMPLING\_RATE\_SENSORS} permission and explain the purpose of high sampling rates to pass the review of Google Play store~\cite{android12, warning}. However, the high sampling rate is a normal-level permission that does not require either installation-time or run-time consent from the users~\cite{high_samp}. Furthermore, recent research has shown that sampling rates lower than 200 Hz are still sufficient to extract a large portion of critical information that can be recognized by emerging machine learning and deep learning algorithms~\cite{bolton2023characterizing, ba2020learning}. As a result, there is still an urgent need for monitoring zero-permission sensor usage.


\subsection{SensorManager Interface}\label{sec:sensor_manager}

The SensorManager class in Android provides the interface for application developers to interact with the hardware of zero-permission sensors. The applications need to register for a sensor event listener with a callback that processes received new sensor readings (\texttt{onSensorChanged}). When registering the listener, the application needs to request a desired sampling rate of the sensor. Android has four pre-defined sampling rate instances, namely  \texttt{SENSOR\_DELAY\_NORMAL/UI/GAME/FASTEST} which have increasing numeric values approximately ranging from 5 Hz to 400 Hz on most smartphones. The actual value of each instance is implemented by smartphone manufacturers and varies with different types of sensors. Application developers can also directly specify the desired sampling rate values (in millisecond sampling intervals) without using these pre-defined values. Similarly, the range of supported sampling rates for each phone model and each sensor could vary.

It is important to note that the actual sampling rate of the sensor readings provided by the Android OS can be different from what the applications request because the OS also needs to balance the bandwidths of different system operations besides passing sensor readings to applications~\cite{monitorsensor}. This enables the detection method of \sysname (Section~\ref{sec:method}).

\section{Methodology \& Design}~\label{sec:threat_model}
Given the emerging problems of nearly unlimited access to zero-permission sensor data, our work proposes a user-space runtime sensor usage monitoring mechanism and design \sysname to support researchers and end users.

\subsection{Threat \& System Model} \label{sec:threat_model}

We assume third-party apps as data collectors want to acquire data from zero-permission sensors without revealing the use of these sensors and these apps can run in either the foreground or background. \sysname takes the defender's role in the form of a user-space trusted application that data providers such as security researchers and end users can run continuously in the background to monitor details of other third-party apps' sensor usage. Given that \sysname's working principle (Section~\ref{sec:method}) allows it to monitor OS-wise sensor usage instead of individual app's usage, we assume that a data provider that wants to associate detected sensor usage to an exact third-party app can ensure only \sysname and this app are running, which can be achieved by terminating or uninstalling other background user-space apps.

\subsection{Sampling-based Sensor Usage Detection} \label{sec:method}

This section introduces the working principle of \sysname and characterizes its capability of detecting various sensor usage.

\subsubsection{Instant Sampling Rate Variation \& Convergence}\label{sec:rules}

\begin{figure}[!t]
	\centering
\includegraphics[width=.4\textwidth]{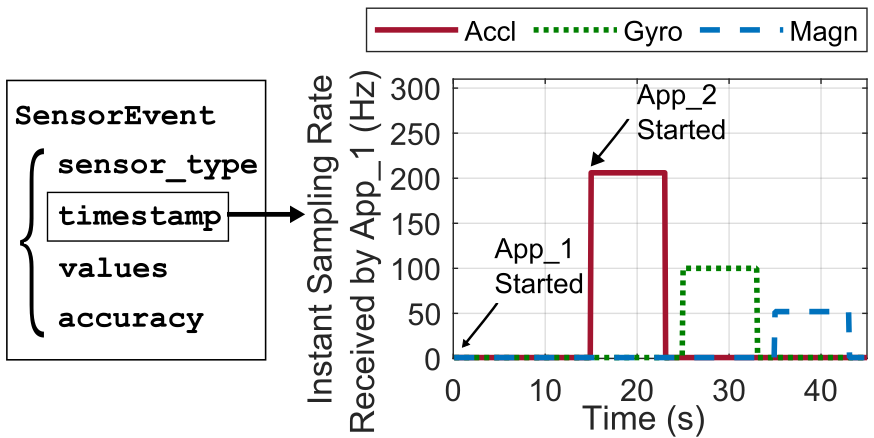}
\vspace{-.1in}
    \caption{The change of instant sampling rates detected by an app (App\_1) when another app (App\_2) starts running and using sensors. \sysname calculates the instant sampling rates using the timestamps of the SensorEvent data packets.} 
	\label{fig:ISR}
    \vspace{-.1in}
\end{figure}

\sysname detects Android sensor usage by observing variations in the actual \textit{instant sampling rate} of sensor readings provided by Android OS, which could be different from the sampling rate  requested by an application (denoted as $f_{req}$). The instant sampling rate at a certain system time $t$ can be calculated as 
\begin{equation} \label{eq:f_{inst}}
    f_{inst}(t) = 1 / (T_a-T_b) 
\end{equation}
where $T_{a}$ and $T_{b}$ stand for the discrete timestamps of the current and last sensor data packet received via the SensorEvent data structure. Our empirical tests with the SensorManager interface show that starting another application could change the instant sampling rates received by a running application that is already collecting sensor data, as shown in Figure~\ref{fig:ISR}.

Based on this phenomenon, we hypothesize that when different applications register to access the same sensor, e.g., the accelerometer on the phone, the instant sampling rates available to each application will be \textit{interdependent}. To verify this hypothesis, we ran several instances of a custom sensor-access application, each requesting a different sampling rate. The test results reveal a uniform sampling rate convergence rule: 
\begin{quote}
    \textit{When multiple applications access the same sensor's data, the instant sampling rate provided to them converges to the highest OS-supported rate requested among all running applications.}
\end{quote}
This implies that a user-space defender application can probe the sensor usage of other applications by checking the instant sampling rate of its own sensor data. Specifically, the defender can register an unusually low sampling rate and observe the increase in its instant sampling rates as a sign of the monitored applications using the same sensor, as demonstrated in Figure~\ref{fig:ISR}. Essentially, this mechanism creates a benign covert communication channel between different applications. 
Our tests find this rule to be independent of when the applications initiate the sensor access requests and whether the applications are running in the background or foreground. We were able to align these observations with the official documentation of Android~\cite{sensorstack},  
\rev{which confirms the convergence to the maximum requested sampling rate but does not specify possible instant sampling rate variation behaviors in different background/foreground scenarios. Our work aims to provide detailed characterizations of this Android's intrinsic but unexplored behavior for building \sysname and collecting the first dataset for evaluating zero-permission sensor usage.}


\subsubsection{Detection Capability}
\label{sec:Detection_Capability}
Our next tests with several self-made sensor data collection apps verify that \sysname could detect zero-permission sensor usage across different smartphone hardware and Android software and is immune to code obfuscations.


\textbf{Device \& Sampling Rate Range.} \sysname is able to monitor sensor usage on Android devices that support the sensor management framework. This includes smartphones, smartwatches, and other Android-powered devices running Android 8 (released in 2017) and all newer versions of Android. Our tests verified the feasibility of using \sysname on six common smartphone models from Google, Samsung, etc., as listed in Table~\ref{tab:phones}. 
\rev{Nevertheless, the condition of observing sampling rates higher than the requested value $f_{req}$ suggests that \sysname cannot detect certain usage smaller than or equal to this value and thus have a limited range of detectable sampling rates. An informed defender will set $f_{req}$ to the minimum supported sampling rate, denoted as $f^i_{min}$ for sensor $i$, ensuring the minium sampling rate is the only usage that is not detected. Different manufacturers and phone models may differ in their implementations of the minimum supported sampling rates, as shown by Table~\ref{tab:phones} for the six phones. While the minimum sampling rate usage will inevitably be overlooked, we hypothesize that this is relatively low likelihood, and further measure to what degree this could affect the detection performance in Section~\ref{sec:eval_performance} and Section~\ref{sec:discussion}.}

\textbf{Code Obfuscation \& Native Codes.} Code obfuscation and native code are popular techniques used by commercial app developers to prevent reverse engineering and achieve better runtime performance. Despite the significant challenges these techniques pose to static analysis methods (Section~\ref{sec:related_defense}), \sysname is found to be immune to these techniques because of its runtime dynamic analysis nature that does not require processing any code/binary files. Our tests with a customized app obfuscated with the ProGuard obfuscator and another customized app that utilizes C/C++ native codes verified that \sysname could detect zero-permission sensor usage in both cases while a previous attempt of adapting existing static taint analysis for detecting sensor data leaks \cite{sun2021characterizing} could not detect these usages. 

\textbf{Web Sensor Usage.} While this work focuses on user-installed third-party Android apps, we find that \sysname is also able to detect zero-permission sensor usage from the web, which has been shown to be another potential threat of sensor data leakage~\cite{das2018sense}. 

\subsection{\sysname Implementation}
\label{sec:system}
\begin{figure}[!t]
	\centering
    \includegraphics[width=0.46\textwidth]{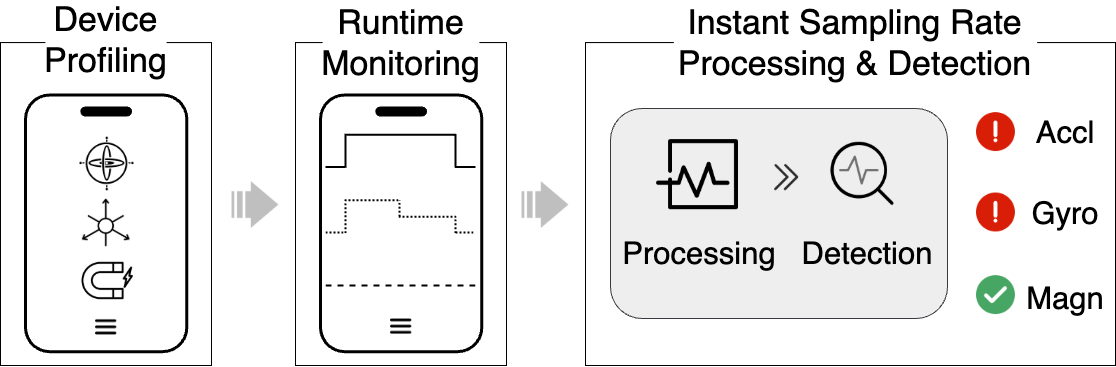}
    \vspace{-.1in}
	\caption{\sysname's workflow.}
	\label{fig:sensoscope_workflow}
    \vspace{-1em}
\end{figure}

Figure~\ref{fig:sensoscope_workflow} summarizes the process of using \sysname to characterize sensor usage on an Android device. The core component is the \sysname app that continuously monitors in the background. The current implementation detects the usage of the three mostly used zero-permission sensors: accelerometers, gyroscopes, and magnetometers. Other sensors can be easily added to accommodate more specialized use cases.

\textbf{Device Profiling.} The profiling stage finds the minimum supported sampling rates $f^i_{min}$, which enables the data provider to set an appropriate threshold for the observed sampling rates $f_{inst}$ to assert the use of each sensor in the detection stage. Each smartphone model with the same software OS version only needs to be profiled once. The \sysname app already implements a profiling mode to automatically achieve this. The profiling mode runs the app in the foreground, requests several sampling rates, and determines $f^i_{min}$ by checking the actual received instant sampling rates. The whole profiling process takes less than 2 min. 

\textbf{Runtime Monitoring.}
Entering the monitoring mode, the \sysname app runs in the background while the data provider opens another app they want to analyze. The tested app can run either in the foreground or background. The \sysname app stores instant sampling rate data, i.e., the readings of the three sensors. 

\textbf{Instant Sampling Rate Processing \& Detection.}
After collecting instant sampling rate data, \sysname calculates and examines a time series $f^i_{inst}(t)$ for each sensor $i$ according to Equation~\ref{eq:f_{inst}}.
Our pilot testing observes that the fluctuation of instant sampling rates sometimes includes a few outliers, as shown in Figure~\ref{fig:outliter_cleaning}. Such outliers are mostly caused by the transitions between different used sampling rates and may cause unstable sampling rates calculations.  
We thus implement a processing step to clean the time series. We define an outlier in $f_{inst}$ as a sampling rate value with fewer than three consecutive occurrences and replace these values with its preceding or following value which is closer to the outlier value's magnitude. 
After cleaning the time series, we find that the time series data remains relatively stable with fluctuation errors always in the range of 0.4 Hz (Figure~\ref{fig:freq_error_bar}). \rev{The range of such fluctuations is affected by factors such as OEM and software variations.}
We thus empirically set the sensor usage detection threshold to $f^i_{min} + 0.5$.
\rev{Following this methodology, the actual thresholds used for different devices can be adjusted after collecting sensor data in the device profiling phase.}
By default, this work declares sensor usage when  
$f^i_{inst}(t) > f^i_{thres} = f^i_{min} + 0.5$.

\section{Evaluation} \label{sec:eval}

Our evaluation aims to explore the following research questions:
\begin{itemize}
    \item \textit{RQ1}: To what degree can \sysname reliably detect zero-permission sensor usage (Section~\ref{sec:eval_performance})?
    \item \textit{RQ2}: How does the overall landscape of real-world sensor usage look like (Section~\ref{sec:eval_overall}-\ref{sec:eval_fgbg})? 
    \item \textit{RQ3}: What specific interesting sensor usage behaviors can \sysname reveal (Section~\ref{sec:eval_cases})?
\end{itemize}

\subsection{Experimental Setup}

\subsubsection{Dataset}
We collected two Android application datasets, totaling 1,448 apps, to answer these questions. A Detection Performance evaluation dataset with 50 apps measures the precision and recall of \sysname in detecting sensor usage; A Sensor Usage dataset with 1,398 popular apps from Google Play allows us to analyze the widespread abuse of zero-permission sensors in real-world scenarios. By default, the evaluation uses the OnePlus Nord N200 phone with Android 12 as it has $f^i_{min}$ of 1 Hz for gyroscopes and magnetometers and 5 Hz for accelerometers, providing a relatively large range of detectable instant sampling rates. Additionally, we also use the Samsung Galaxy S9 phone with Android 10 to analyze how different versions of Android, such as the introduction of the \texttt{HIGH\_SAMPLING\_RATE\_SENSORS} permission in Android 12, could affect the pattern of sensor usage. The datasets thus contain APK/XAPK files for the two devices. 

\textbf{Detection Performance.} Since there currently does not exist a ground-truth dataset for Android sensor usage, we constructed such a dataset consisting of 50 apps that are known to use sensors. We collected these 50 apps from Google Play by manually searching for apps that certainly use zero-permission sensors, such as those named (1) compass,  magnetometers, EMF/metal detectors, etc., (2) acceleration meters, vibration meters, speedometers, etc., and (3) gyroscopes. These three categories of apps are known to use at least the phones' magnetometer, accelerometer, and gyroscope.

\begin{table*}[!ht]
\centering
\caption{Detected Sensor Usage of 1398 Most Popular Google Play Applications}
\vspace{-.15in}
\label{tab:popular}
\begin{tabular}{c|c|c|c|c|c|c}
\hline
\textbf{Device} & \textbf{Android Version}  & \textbf{Accelerometer} & \textbf{Gyroscope} & \textbf{Magnetometer} & \textbf{Any} &\textbf{All Three} \\ \hline\hline

Samsung Galaxy S9 & 10 (API 29) & 639 (45.7\%) & 475 (34.0\%) & 359 (25.7\%) & 645 (46.1\%) & 310 (22.2\%) \\ \hline
OnePlus Nord N200 5G & 12 (API 31) & 557 (39.8\%) & 485 (34.7\%) & 320 (22.9\%) & 583 (41.7\%) & 300 (21.5\%) \\ 
\hline
\end{tabular}
\end{table*}

\textbf{Sensor Usage.} The dataset consists of 1398 most popular Google Play applications in the U.S. ranked by Appfigures~\cite{Appfigures}. APK/XAPK files are downloaded from APKPure and APKCombo. These apps span 35 different categories to ensure a broad representation of sensor usage patterns. The popularity and diversity of these applications aim to provide valuable insights into sensor access trends and potential abuse risks.


\subsubsection{Procedure.} \label{sec:eval_procedure}
The experiment tests one app in the dataset at a time, whose operations including installation, launching, etc., are automated by Python scripts running on a MacBook that utilizes the Android Debug Bridge (ADB) and the uiautomator2 tool~\cite{uiautomator2}. The procedure of testing each app includes the following steps: (1) \sysname starts recording; it first runs for 5 seconds to ensure the recorded instant sampling rates equal $f^i_{min}$ so as to confirm that no other software was accessing the sensors during the test. (2) The tested app starts running in the foreground for 15 seconds to examine its foreground sensor usage. (3) The tested app is then put in the background and runs for another 15 seconds. (4) The tested app is stopped and uninstalled and \sysname stops recording.  

By default, the testing does not apply UI interactions to the tested apps. Since specific UI interactions and app events may trigger additional sensor usage, as will be shown in Section~\ref{sec:eval_cases}, this testing procedure provides a lower-bound baseline for the number of apps using sensors. Additionally, we captured periodic screenshots of the tested app using ADB to manually verify whether the app was successfully installed and launched.

\subsection{Detection Performance \& Overhead} \label{sec:eval_performance}
This section evaluates \sysname's output when there is known sensor usage and the runtime overhead \sysname incurs.

\subsubsection{Using Sensors} 
Our tests show that \sysname could successfully detect all sensor usage of the 50 apps. Among these apps, the minimum detected sampling rate of the used sensors is about 50 Hz, which corresponds to the constant values set by \texttt{SENSOR\_DELAY\_GAME}.  The results provide evidence that \sysname can detect zero-permission sensor usage of most commercial apps with very high probabilities because apps are not likely to use $f^i_{min}$ given the overly limited amount of information $f^i_{min}$ can provide. \rev{A Frida-based dynamic analysis that aims to provide baseline measurements for commercial applications on a rooted phone will further confirm this observation (see Section~~\ref{sec:discussion}).}

\subsubsection{Not Using Sensors} \label{ref:no_sensors}
To further verify that \sysname does not cause false positives of sensor usage detection, we perform two additional tests. First, we ran \sysname in the background without any foreground apps to examine whether the Android OS activities could trigger sensor usage. Second, we ran a simple self-made application that does not use sensors to examine whether non-sensor app activities could trigger sensor usage.  We ran both tests for 30 min continuously and 10 times each throughout the day, which provides a prolonged time frame that facilitates the assessment of long-term behaviors. The screen auto-rotation feature of the device was turned on in these tests, where sensors are used by the OS to detect change of phone orientations. \sysname did not mistakenly report any sensor usage in these cases. This confirms that changes in the instant sampling rates occur only when the SensorManager interface is explicitly invoked in user space.

\subsubsection{Power \& CPU Overhead}
\label{sec:eval_overhead}

In the above tests of no foreground apps, we simultaneously profiled \sysname's CPU and memory usage using AccuBattery~\cite{accubatery} and Android Studio's built-in Profiler. 
According to AccuBattery, running \sysname for background monitoring increased the battery usage from 2.0\% of its total battery capacity per hour to 5.6\% per hour, incurring a power consumption overhead that consumes approximately 2.6\% of the battery per hour. This overhead is comparable to that induced by common background monitoring applications such as AccuBattery itself. The Profiler determined that CPU usage is light throughout \sysname's runtime out of the three categories of "Light", "Moderate", and "High". The relatively light power and CPU usage overheads are due to 
the fact that \sysname only needs to use a sampling rate of as low as 1 Hz to access sensors in the background. This makes it feasible to use \sysname not only for security research that does not face any limits of overheads but also for everyday background monitoring by average phone users who might care about overheads.

\subsection{Popular Google Play Applications}
\label{sec:eval_overall}
Table~\ref{table:percentages} summarizes the percentage of apps using each sensor, showing zero-permission sensor usage is very common both before and after Android's change in zero-permission sensor regulation in Android 12. We additionally analyzed 150 apps manually that were randomly selected from the dataset and found that \textit{none of these apps explicitly stated their purposes for using zero-permission sensors}, either in the app UI or Google Play webpage introduction.

On the Samsung Galaxy S9 device with Android Android 10, 46.1\% of the 1398 apps use at least one of the three sensors. The accelerometer is used the most, followed by the gyroscope, with the magnetometer being the least used (25.7\%). 
The pattern of sensor usage on the OnePlus Nord N200 device with Android 12 remains consistent with slight decreases in accelerometer and magnetometer usage and a slight increase in gyroscope usage. Given that the Samsung Galaxy S9 and OnePlus Nord N200 devices have different $f^{accel}_{min}$ which result in 5.5 Hz and 1.5 Hz accelerometer usage detection thresholds respectively, we also tested setting $f^{accel}_{thrs}$ of the Samsung device to 5.5 Hz to control the factor. This produces an accelerometer usage percentage of 42.3\% which is 3.4\% lower than using a 1.5 Hz threshold, showing that a small portion of apps used low accelerometer sampling rates in the range 1.5 to 5.5 Hz.  It is also an observed common pattern that the accelerometer, gyroscope, and magnetometer are often accessed together. More than 20\% of the 1398 apps, i.e., over 50\% of all apps using sensors, are found to be using all three sensors simultaneously.

\begin{figure*}[!th]
	\centering
\includegraphics[width=0.99\textwidth]{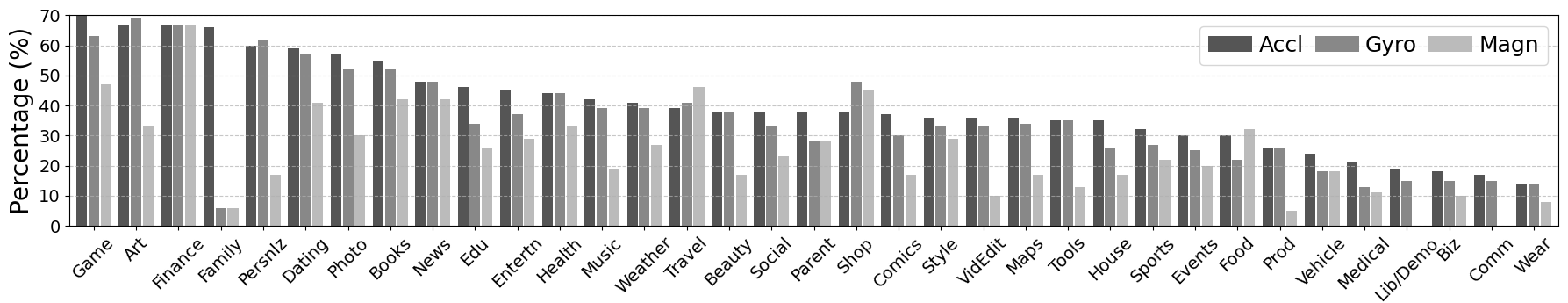}
\vspace{-.2in}
	\caption{Zero-permission sensor usage by different categories.} 
	\label{fig:category}
    \vspace{-1em}
\end{figure*}

\subsubsection{App Categories}
Figure~\ref{fig:category} further provides a breakdown of these usages by app categories. Table~\ref{table:percentages} provides more detailed data with several representative apps in each category. While most categories of apps follow the pattern that the accelerometer and magnetometer are the most and least used sensors among the three, there are some exceptions. For example, the categories Art/Design and Personalization use gyroscope the most. In addition, Shopping and Travel/Local use magnetometers more than gyroscopes and accelerometers, which is likely due to localization services these apps provide that depend on magnetometer data for navigation. 

The most outstanding problem observed is the unclear and unexplained purpose of sensor usage in many app categories.  
Surprisingly, categories that people often do not expect to have apparent dependencies on sensor-collected information, such as apps in Finance, Dating, Book/Reference, Shopping, Weather, etc., actually heavily depend on accelerometer and gyroscope data for unspecified purposes. 
For example, more than half of the Book/Reference apps use accelerometer and gyroscope data but identifying the rationale behind this usage is challenging. Furthermore, although the results of certain categories such as Health/Fitness seem to align with expected sensor usage, a closer look at the individual apps raises questions, such as why the subcategory of exercise planning apps that are not supposed to monitor user motions still frequently collect sensor data. 

In addition, results such as the sensor usage by 67\% of the Finance apps collecting data from all three sensors reveals that real-world applications might already be using sensor data for identification. Our literature review shows that Finance apps use sensor data information for authenticating and tracking users and devices~\cite{PayPal_Magnes, Appsflyer}. While the collected sensor data is theoretically used for benign purposes of enhancing security in these cases, it also echoes the findings of prior research that real-world applications have the capability of exploiting sensor data for sensitive information~\cite{PayPal_Magnes_leak}.

\subsection{Sampling Rate Distribution}
\label{sec:eval_distribution}

This section analyzes the sensor sampling rates used by popular Google Play apps and the impact of high sampling rate permission introduced by Android 12. 

\begin{figure}[!t]
	\centering
    \includegraphics[width=.4\textwidth]{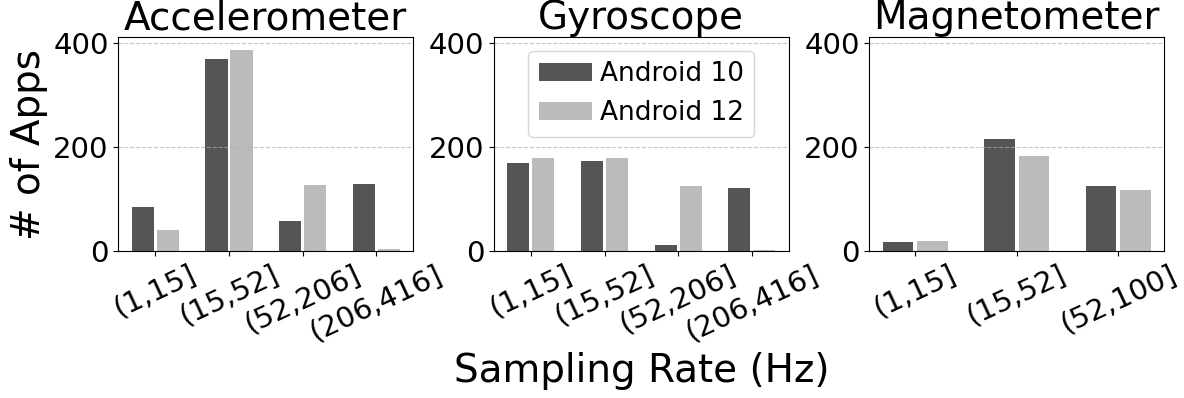}
    \vspace{-.1in}
	\caption{Distribution of Google Play apps' zero-permission sensor sampling rates before and after Android 12.}
\label{fig:iso_max_freq_distribution}
\vspace{-1em}

\end{figure}

Figure~\ref{fig:iso_max_freq_distribution} shows the histogram of the maximum detected sampling rates of each app for the three sensors in the two Android versions respectively. We observe that while many of the app developers tend to request sampling rates defined by the Android-defined constants \texttt{SENSOR\_DELAY\_FASTEST/GAME/UI/NORMAL}, which correspond to sampling rates around 400, 50, 15, and 5 Hz, there are many apps manually requesting other sampling rates such as 200 and 100 Hz.  

The highest sampling rate defined by \texttt{SENSOR\_DELAY\_FASTEST} corresponds to 416 Hz for the accelerometer and the gyroscope and 100 Hz for the magnetometer of the two phones. In Android 10, 20.0\%, 25.5\%, and 34.5\% apps used these highest sampling rates for the three sensors respectively, with 120 out of 130 (92.3\%) applications simultaneously accessing the highest frequency for all
three sensors. These numbers dropped to 0.8\%, 0.4\%, and 32.0\% in Android 12. Out of the 128 apps that used \texttt{SENSOR\_DELAY\_FASTEST} for accelerometer and gyroscope in Android 10, only 5 of them took the effort of declaring the high sampling rate permission in Android 12 to keep using it. On the contrary, the \texttt{SENSOR\_DELAY\_FASTEST} usage for the magnetometer only observed a negligible decrease because 100 Hz is not regulated by the 
\texttt{HIGH\_SAMPLING\_RATE\_SENSORS} permission. The results indicate that the regulation introduced by Android 12 effectively limited certain unnecessary high sampling rate usage. 

We found that \texttt{SENSOR\_DELAY\_GAME} responds to a sampling rate of 52 Hz for accelerometers and gyroscopes in Android 12. Interestingly, however, the highest sampling rate apps can get without declaring the \texttt{HIGH\_SAMPLING\_RATE\_SENSORS} permission in Android 12 is 206 Hz on the OnePlus device and about 21.0\% of apps used this sampling rate. Requesting sampling rates higher than 206 Hz without declaring the \texttt{HIGH\_SAMPLING\_RATE\_SENSORS} permission in Android 12 and above results in compiling errors. This phenomenon suggests two insights. First, many apps tend to use the highest available sampling rates without declaring the \texttt{HIGH\_SAMPLING\_RATE\_SENSORS} permission even if they have to manually test and specify feasible sampling rate values. Second, while Android 12 and above officially specify that the maximum available sampling rates without declaring \texttt{HIGH\_SAMPLING\_RATE\_SENSORS} should be no more than 200 Hz~\cite{android200Hz}, the implementation on the OnePlus device of 206 Hz sampling rates suggests \textit{the potential for manufacturers to violate or bypass this requirement. It remains unclear how Android could enforce this requirement to regulate manufacturers.}

\subsection{Foreground \& Background Sensor Usage}
\label{sec:eval_fgbg}

Our results show that apps tend to have distinct foreground and background sensor usage patterns, especially across different app categories. Out of the 583 apps that used sensors, all of them had foreground access while 168 of them (28.8\%) had background access. 

\subsubsection{Foreground Usage.} 
\begin{figure}[!t]
	\centering
    \includegraphics[width=.45\textwidth]{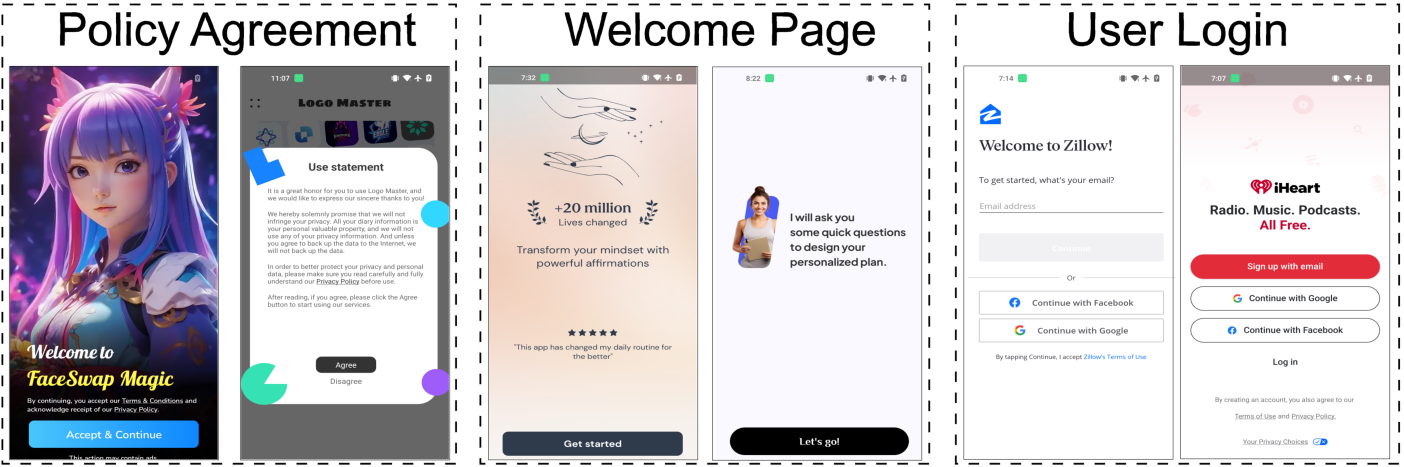}
    \vspace{-.1in}
	\caption{Examples of unexpected zero-permission sensor usage when apps start in the foreground.}
	\label{fig:WelcomePage}
    \vspace{-.15in}
\end{figure}

Although, at first glance, the high percentage of foreground sensor usage might be thought of as being associated with user interactions, our experiments already eliminated the impact of this factor by not having any UI inputs in the testing procedure as mentioned in Section~\ref{sec:eval_procedure}. 
Thus, the sensor data collections occured without any user interaction after opening the application. For example, Figure~\ref{fig:WelcomePage} shows three typical types of foreground activities after the apps start that are not supposed to have reasonable sensor-related operations, such as asking users to agree to the apps' policies, showing a welcome message, and prompting users to log in. Notably, we found that about 65\% of the apps with foreground sensor access paused at these pages.

\subsubsection{Background Usage}
The highest background sensor usage ratios were observed in Productivity (60.0\%), Communication (57.1\%), and Games (51.4\%). While Games apps such as racing are expected to heavily depend on sensor data for user hand gesture tracking, the majority of their background usage appears unexplained as any background activities do not change the states of the apps. Such background usage is thus likely to be negligent sensor abuse due to poor mobile device resource management practices. The high background usage ratios in Productivity and Communication appear less intuitive. On the contrary, the Android-Wear category that provides services to resource-constrained wearable devices demonstrated no (0\%) background sensor usage. This suggests that these applications, at a minimum, require explicit user interaction to activate sensors rather than maintaining passive access in the background, making them a good example of responsible sensor usage in background stages.

\subsubsection{Foreground-Background Transition} 
\begin{figure}[!t]
	\centering
    \includegraphics[width=0.48\textwidth]{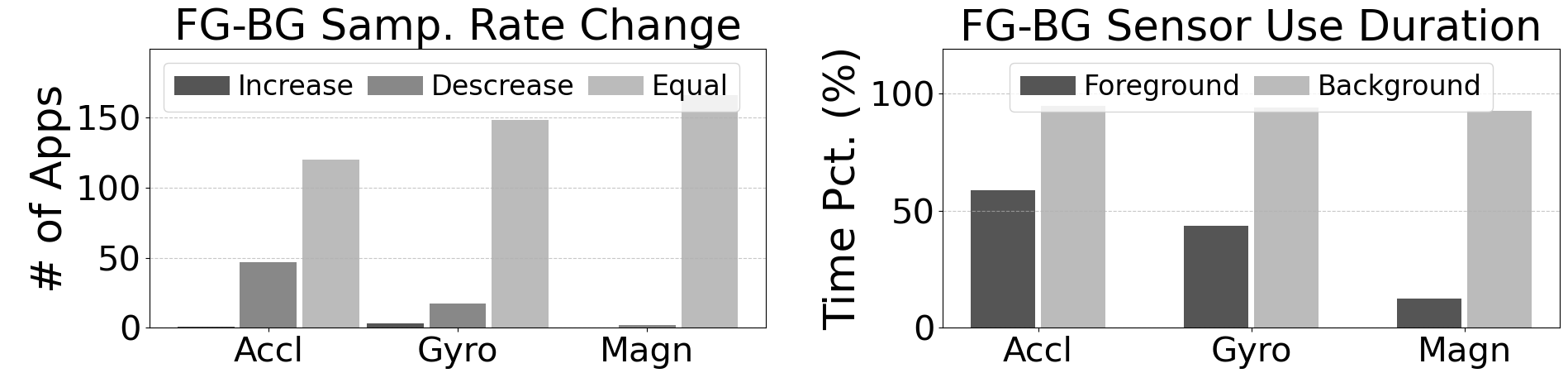}
    \vspace{-.2in}
	\caption{Background sensor usage tends to have decreasing or equal sampling rates (left) but higher durations (right).}
	\label{fig:fg_bg_diff_show}
    \vspace{-1em}
\end{figure}

Comparing sensor usage in the foreground and background shows that apps generally use lower sensor sampling rates in the background, while simultaneously extending the duration of sensor access, as shown in Figure~\ref{fig:fg_bg_diff_show}. The three types of zero-permission sensors exhibit significantly higher peak access sampling rates in the foreground, with a wide interquartile range and numerous high sampling rate outliers. Once transitioned to the background, the maximum frequency drops substantially.  This verifies apps' common behaviors of shifting toward persistent, long-term monitoring in the background without user awareness.




\subsection{Case Study} \label{sec:eval_cases}
This section further discusses several use cases of \sysname that revealed interesting sensor usage behaviors during our examination of individual apps.

\textbf{Persistent Sensor Access After App Termination.} It was surprising to observe that certain apps could continue to access zero-permission sensors after being killed in GUI or even force-stopped using ADB.  
More than 50 (3.6\%) apps exhibited \rev{reproducible} persistent sensor activity after termination \rev{on the OnePlus phone}. 
\rev{
However, running the same apps on a Pixel 3 smartphone did not show the same persistent sensor usage. We believe this OEM variation is caused by the fact that Android allows manufacturers to customize its behaviors of process life-cycle and background service handling~\cite{android_bg}. Additionally, other factors such as specific sensor hardware and drivers could also cause different sensor access termination behaviors. 
This indicates further challenges of enforcing strict sensor access policies across different platforms.}  

\color{revcol}

\textbf{Extended Running and User-triggered Sensor Usage.} The evaluations so far ran each app for only 15 seconds in the foreground without any user interactions. This setting provides a lower-bound measurement, indicating that more extensive sensor data collection and abuse could exist. To confirm this, we manually tested 35 randomly selected apps, one from each category. 18 of them had foreground sensor usage detected in previous tests. We had normal GUI interactions with each of these apps for 5 minutes continuously. 
The tests found additional sensor usage triggered by user interactions in 14 out of these 18 apps. The most common trigger is user logins (6 apps), such as hitting the ``Continue With Google'' button. This suggests the sensor data is used as an information source for authentication. Other triggers also include accepting policies, granting permissions, and other app-specific actions.
Furthermore, user interactions triggered sensor usage in 5 of the 17 apps that did not use sensors in the 15-second foreground tests. This suggests future research could integrate \sysname with automated GUI testing frameworks for large-scale measreument. 

\color{black}

\begin{figure}[!t]
	\centering
    \includegraphics[width=.48\textwidth]{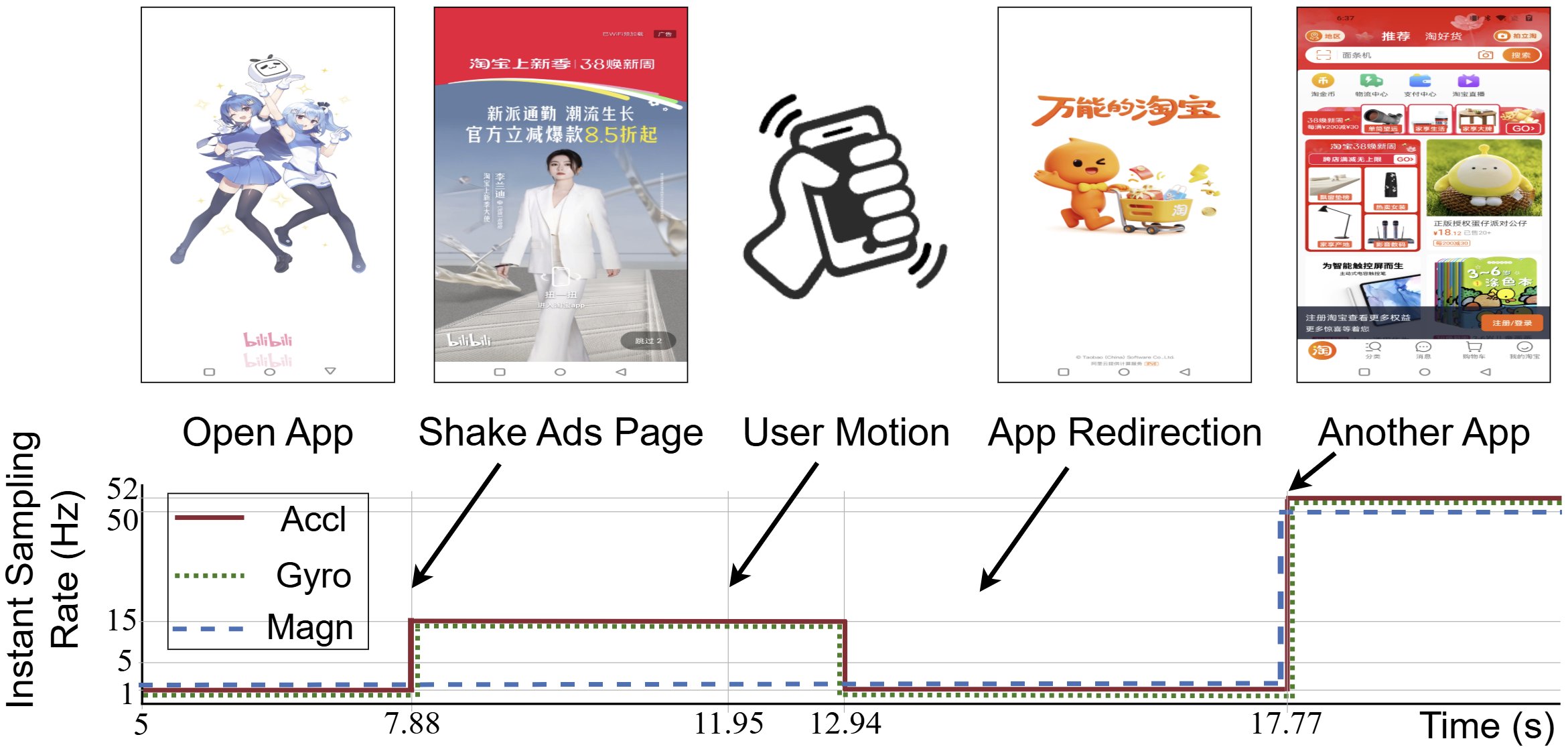}
    \vspace{-.2in}
	\caption{\sysname detects the sensor usage behind emerging shake-ad features of some apps.}
	\label{fig:ShakeJump}
    \vspace{-1em}
\end{figure}

Furthermore, \sysname's capability of monitoring time-variant sensor usage correlated with user interactions further enables analysis of sensor-based exploitation including \textit{shake-triggered advertisement} found on apps such as Bilibili--one of the most popular video sharing platforms. After a brief opening page, the app enters a shake-ad activity that continuously collects sensor data for 5 seconds. The shake-ad page has a concealed prompt of ``shake the phone to enter Taobao''. Even when users did not notice this prompt, users' minor involuntary hand movements could be detected by the accelerometer and gyroscope with sampling rates of 15 Hz (\texttt{SENSOR\_DELAY\_UI}) and be regarded as users actively shaking their phones, thus redirecting to the Taobao shopping app by Alibaba (Figure~\ref{fig:ShakeJump}). The Taobao app continues to collect data from the two sensors for unknown purposes, at 52 Hz (\texttt{SENSOR\_DELAY\_GAME}). This discovered pattern confirms what has been hypothesized in previous research such as triggering a malicious app via sensors~\cite{sikder20176thsense}.

\color{revcol}

\textbf{Third-party Libraries.} The observed common sensor access patterns in different applications suggest a hypothesis that some zero-permission sensor usage could be caused by shared third-party libraries. 
To provide preliminary insights, we decompiled representative apps using jadx~\cite{jadx} and analyzed their obfuscated source codes. 
We found a large portion of apps with the unique pattern of accessing sensors for a very short time at app startup, with the highest possible sampling rates. Reading their decompiled code could accurately identify this sensor usage by Appsflyer~\cite{Appsflyer}, a marketing analytics service that helps app developers track and optimize user acquisition campaigns. Appsflyer records data from the three sensors for 500 ms, and produce a hash from the saved data to match users/devices. Our dataset shows that about 49\% of the apps using sensors have included the Appsflyer library. 
In contrast, PayPal Magnes~\cite{PayPal_Magnes}, which is a widely used mobile payment SDK for collecting real-time device data to help detect fraud, continuously saves collected sensor data at low sampling rates (about 20 Hz) to JSON files during its operation, and then sends the files to its remote servers. About 6\% of the apps using sensors included code of PayPal Magnes service. Additionally, the sensor access patterns of multiple libraries may be superimposed. 
These results also suggest possible follow-up research of identifying third-party libraries by analyzing sensor behaviors detected by \sysname from user space.

\color{black}

\section{Discussion} ~\label{sec:discussion}

This section discusses the limitations and possible future research directions revealed by \sysname.

\color{revcol}

\textbf{Sensor Usage Ground Truth \& Baseline.} As Section~\ref{sec:eval} explained, the lack of a ground truth presents a major roadblock for zero-permission sensor monitoring. 
The datasets collected by our work take the first step toward bridging this gap, but unavoidably faces limitations of lacking a baseline that \sysname could compare with. 
In particular, certain apps could still be using the minimum supported sampling rates $f^i_{min}$ and thus bypass the detection of \sysname. 
Aiming to probe how to provide such a baseline, our work explored applying the Frida dynamic instrumentation toolkit~\cite{frida} on a rooted smartphone to monitor the invocation of the \texttt{onSensorChanged} callbacks. The results show that out of the 830 apps for which \sysname did not detect zero-permission sensor usage, only 38 (4.49\%), 3 (0.35\%), and 10 (1.18\%) of them use the accelerometer, gyroscope, and magnetometer at their respective $f^i_{min}$. This confirms that commercial applications mostly use non-minimum sampling rates to collect more sensor data. The higher percentage of accelerometer is due to a relatively high $f^{accl}_{min} = 5$. 
To further reduce the risks of undetected sensor usage at the minimum sampling rates, we recommend that privacy-aware manufacturers implement $f^i_{min} = 1$ for all sensors on future devices. It is also worth noting that Frida only provides a baseline for comparison instead of the ultimate ground truth due to the possible Runtime Android Self Protection measures (RASP) that may prevent dynamic instrumentation or running apps on rooted devices. This motivates future works of comprehensive ground truth development.






\color{black}

\rev{\textbf{Sensor Usage Attribution.} Our results so far reveal important follow-up questions, such as how to distinguish between benign and malicious sensor access patterns. Determining whether a zero-permission sensor is exploited to maliciously acquire private information that users do not want to share poses significant new challenges compared to prior research on identifying malicious Android applications. This is because, unlike IMEI or other textual data targeted by conventional malicious apps that can be directly associated with entities of private information after being detected in the saved files or network packets, the captured raw sensor data needs to be processed by the sensor-based malicious apps themselves to extract semantic signals that correlate with private information (Section~\ref{sec:related_exploitation}). Thus, identifying a sensor-based malicious app is equivalent to completely reverse engineering the app's sensor-related functionalities. 
This challenge could be further amplified when the apps use a server-side exploitation model, where only raw sensor packets are transmitted to a remote server for black-box processing.} These challenges motivate dedicated future research that integrates reverse engineering, automated large-scale static analysis, root-based dynamic analysis, and user studies. \rev{In the extremely challenging case of server-side black-box processing, we believe that advanced methods such as differential behavior analysis~\cite{continella2017obfuscation} need to be developed to empirically probe whether collected sensor data is used for malicious or benign purposes.} 
Although these topics fall out of the scope of this work, the sensor usage patterns measured by \sysname could serve as motivating examples and baselines for future research. 

\textbf{Environments \& Factors.} Our evaluations demonstrated the effectiveness of \sysname in lab environments with a relatively small number of varying factors. 
\rev{Our measurement analysis focused on Android 12, whose sensor-related policies have remained consistent up to Android 15 at the time of writing. Without further policy changes, running \sysname on other devices will not be affected by the newer Android versions. Instead, we found that specific OEM configurations, such as different minimum supported sampling rates (e.g., Table~\ref{tab:phones}), could have larger impact on \sysname's results. We believe this encourages future research by dedicated security analysts that utilizes \sysname to perform in-depth app or device-wise analysis. Potentially, an online community and database could be created to crowdsource \sysname's measurements on various devices.}

\textbf{Future Sensor Privacy Enhancement.}
While the current \sysname app can be readily used by lay users for sensor usage notification (as shown in Figure~\ref{fig:sensor_framework}), additions to the app's UI functionality
will further improve its usability as a privacy
enhancement tool. 
\rev{A limitation of the current \sysname design is the requirement of running a single app besides the monitor. In daily monitoring scenarios, detection performance could degrade if multiple third-party apps run concurrently because \sysname can only ascertain that at least one app is using sensors.  
It is possible to address this with more sophisticated \sysname designs, which could require additional permissions, such as \texttt{UsageStats}, to monitor when each app is active in the foreground/background to create a correlation engine that maps temporal patterns of sensor usage with app activities. As a user-trusted app, \sysname may utilize Android's accessibility feature~\cite{accessibility} to monitor GUI activities that trigger sensor usage and thus pre-build a signature database for different apps' sensor access patterns.} 

\color{revcol}

Our findings also motivate the need for enhanced privacy protections after detecting zero-permission sensor abuses. On the policy level, we suggest platform providers such as Google Play implement knowledge-based auditing and stricter review processes to regulate unnecessary sensor access. This could require adding explanations of sensor data usage, as a standard procedure for generating SBOM~\cite{camp2021sbom} for Android and similar ecosystems. 
We have disclosed the identified sensor abuse problems to Google and aim to provide detailed mitigation suggestions in further interactions. 
From a technical perspective, future research could investigate how to filter out sensitive information or selectively inject controlled noise into sensor data streams to preserve utility while mitigating privacy leakage~\cite{bolton2020curtail}. 
Although some related works (Section~\ref{sec:related_defense}) explored similar approaches, finding more deployable and scalable mitigation techniques that may not require OS modifications or root privileges remains an open challenge. More fine-grained permission models for sensor data are needed, though helping users understand privacy implications without adding cognitive burden remains challenging~\cite{benisch2011capturing}. This is further complicated by sensor usage attribution difficulties identified in this work. We see significant opportunities for research into effective and usable privacy-preserving sensing mechanisms.


\color{black}


\section{Related Work}
This section introduces prior research investigating the threats of zero-permission sensor usage and the possible ways of mitigation.

\subsection{Zero-permission Sensor Exploitation}\label{sec:related_exploitation}

Research has extensively shown threats posed by zero-permission sensors. \cite{sikder2021survey} provides a survey that summarizes these existing problems in a systematic way. 
In 2014, Gyrophone~\cite{michalevsky2014gyrophone} first analyzed how sound generated by electronic speakers in the vicinity of smartphones can be captured by phone applications accessing the phones' gyroscope readings.  A series of follow-up research explored the possibility of using accelerometers for speech eavesdropping~\cite{anand2019spearphone, ba2020learning, hu2022accear}.
Similarly, users inputting texts or PIN on the phone touchscreen or walking with the phone cause unique and discernible phone motions that can be captured by zero-permission motion sensors~\cite{cai2011touchlogger,owusu2012accessory,lee2017implicit,long2022side}. Magnetometers are another type of the most analyzed zero-permission sensors for leaking information. Several recent works discovered that smartphones' electromagnetic emissions can be received by the built-in magnetic sensors, which contain signals that enable applications to infer specific phone CPU and display activities~\cite{cheng2019magattack}, or device locations~\cite{block2018my}.  
\rev{While these prior works verified the privacy impact of malicious zero-permission sensor exploitation, it remains unknown to what degree commercial applications have been using data from these sensors and whether the measured usage patterns could reveal potential abusive problems~\cite{sikder20176thsense}. 
This work seeks to address this gap by developing \sysname that measure the first collected dataset of zero-permission sensor usage detection.}

\subsection{Android Sensor Analysis \& Protections}\label{sec:related_defense}

Analyzing and protecting against the possible exploitation of Android sensors is an emerging research field motivated by the observed zero-permission sensor problems. 

\textbf{Static Analysis-based Usage Detection.} Although there has been a large body of taint-analysis such as FlowDroid~\cite{arzt2014flowdroid}, these early works did not consider sensors as an input source that could leak information.  Liu et al.~\cite{liu2018discovering} developed a tool SDFDroid, which disassembles APK files to smali code files using Apktool and looks for sensor listener registration and data receiving callbacks. SDFDroid has the limitations of not performing well on applications with code obfuscation and native code, and cannot detect which exact sensors are used. A later work by Sun et al.~\cite{sun2021characterizing} extends FlowDroid to detect leaks with sensor data sources. Despite its efforts to develop a rule-based approach to identify the sensor types, its approach is still limited by obfuscations and native codes, and only works in the case of a single sensor type being registered. These gaps show the common limitations of static analysis-based methods for detecting Android sensor usage. Furthermore, these works' aims also differ from this research. While they focus on developer-side analysis which requires sophisticated steps of building development environments and code dissembling, our work aims to provide a ready-to-use tool that can be utilized by any user and researcher. 

\textbf{OS and App Instrumentation.} Some other works seek to provide better security against zero-permission sensors by directly modifying the existing Android OS or third-party applications. 6thSense~\cite{sikder20176thsense} developed a context-aware sensor-based attack detection framework that monitors sensor data, infers the current use context, and then decides whether the current sensor use might be malicious. To detect the use context, 6thSense needs to acquire all sensor readings sent to each application, which requires 6thSense to be built into the operating system. 
Sensor Guardian~\cite{bai2017sensor} and SemaDroid~\cite{xu2015semadroid}, on the other hand, choose to directly control applications’ access to sensors by inserting hooks into applications' Dalvik byte-codes or modifying OS implementations to enforce additional control policies at runtime. This requires developers to build a sophisticated development chain in order to modify and control Apps and is limited to use by specialized experts. Furthermore, this approach does not work with apps implementing RASP techniques. In contrast to these approaches, \sysname focuses on the user space sensor usage detection without modifications to the existing Android OS or requiring root privilege. This provides complementary protection capabilities when root and expert privileges are not available.

\section{Conclusion}

This work investigated monitoring zero-permission sensor usage on Android and developed \sysname, a user-space tool allowing researchers and users to analyze when, which, and at what sampling rates third-party apps access these sensors. Our evaluation with 1,448 commercial apps revealed widespread, unjustified sensor usage across multiple app categories, highlighting risks from unregulated access and limitations in current Android security policies. Our findings call for improved regulations and advanced privacy tools to address growing threats. We present \sysname and our dataset as an important step toward mitigating the information asymmetry between data collectors and providers.

\begin{acks}

We appreciate the insights and remarks from our reviewers. This research was supported in part by the National Science Foundation (NSF) Industry-University Cooperative Research Centers Program, CHEST, under grant IUCRC-1916762, and by the NSF award 2330264. Any opinions, findings, conclusions, or recommendations expressed are those of the authors and not necessarily of the NSF.

\end{acks}


\bibliographystyle{ACM-Reference-Format}
\bibliography{texts/refs}

\appendix

\section{Appendix: Supplementary Materials}

\vspace{-.1in}

\begin{table}[!h]
\footnotesize
\caption{Examples of Possible Information Leakage}
\vspace{-.15in}
\label{tab:sensor_table}
\begin{tabular}{c|c|c}
\hline
\textbf{Category} & \textbf{Sensor} & \textbf{Info. Leakage Example} \\ \hline\hline 
\multirow{6}{*}{Motion} & Gyroscope & Speech audio~\cite{michalevsky2014gyrophone, sun2023stealthyimu} \\ \cline{2-3} 
 & Gyroscope & Lock information~\cite{maiti2018towards} \\ \cline{2-3} 
 & Accelerometer & Speech reconstruction~\cite{hu2022accear, ba2020learning} \\ \cline{2-3} 
 & Accelerometer & Touchscreen input~\cite{owusu2012accessory, berend2018there} \\ \cline{2-3} 
 & Accl. + Gyro. & User age \& identity~\cite{davarci2017age, lee2017implicit} \\ \cline{2-3} 
 & Accl. + Gyro. & Keystroke~\cite{cai2011touchlogger, wu2023privacy} \\ \hline
\multirow{3}{*}{Position} & Magnetometer & User location~\cite{block2018my, narain2016inferring} \\ \cline{2-3} 
 & Magnetometer & App activity~\cite{cheng2019magattack, mohamed2023istelan} \\ \hline

\end{tabular}
\end{table}

\begin{table}[!th]

\captionsetup{justification=centering}
\caption{Examples of Android Phones' Parameters}
\footnotesize
\vspace{-.15in}
\label{tab:phones}
\begin{tabular}{c|c|c|c|c} 
\hline
\begin{tabular}[c]{@{}c@{}}\textbf{Phone~}\\\textbf{Model}\end{tabular} & \begin{tabular}[c]{@{}c@{}}\textbf{Android}\\\textbf{Version}\end{tabular} & \textbf{$f^{accl}_{min}$} & \textbf{$f^{gyro}_{min}$} & \textbf{$f^{magn}_{min}$} \\ 
\hhline{=|=|=|=|=}
Google Pixel 3 & 12 & 5 & 1 & 1 \\ 
\hline
Google Pixel 5 & 12 &  5 & 1 & 1 \\ 
\hline
Google Pixel 6 & 13 & 7  & 2 & 1 \\ 
\hline
OnePlus Nord N200 & 12 & 5 & 1 & 1 \\ 
\hline
Samsung Galaxy S9 & 10 & 1 & 1 & 1 \\ 
\hline
Samsung Galaxy S20 & 13 & 1 & 1 & 1 \\
\hline
\end{tabular}\\
\vspace{.03in}
*The unit of minimum supported sampling rates is Hz.
\vspace{-.25in}
\end{table}

\begin{figure}[!th]
	\centering
    \includegraphics[width=0.44\textwidth]{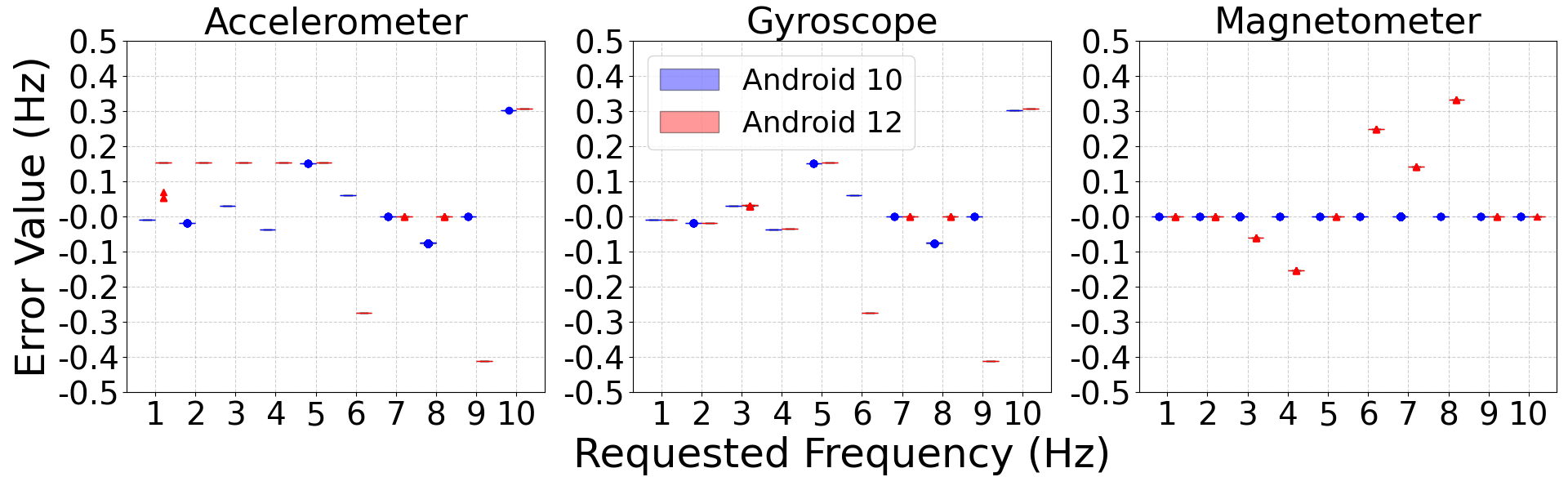}
    \vspace{-.15in}
	\caption{Variation ranges of received instant sampling rates when apps request different sampling rates. The sampling rates remain stable with errors smaller than 0.4 Hz.}
	\label{fig:freq_error_bar}
    \vspace{-.2in}
\end{figure}

\begin{figure}[!th]
  \centering
  \includegraphics[width=0.38\textwidth]{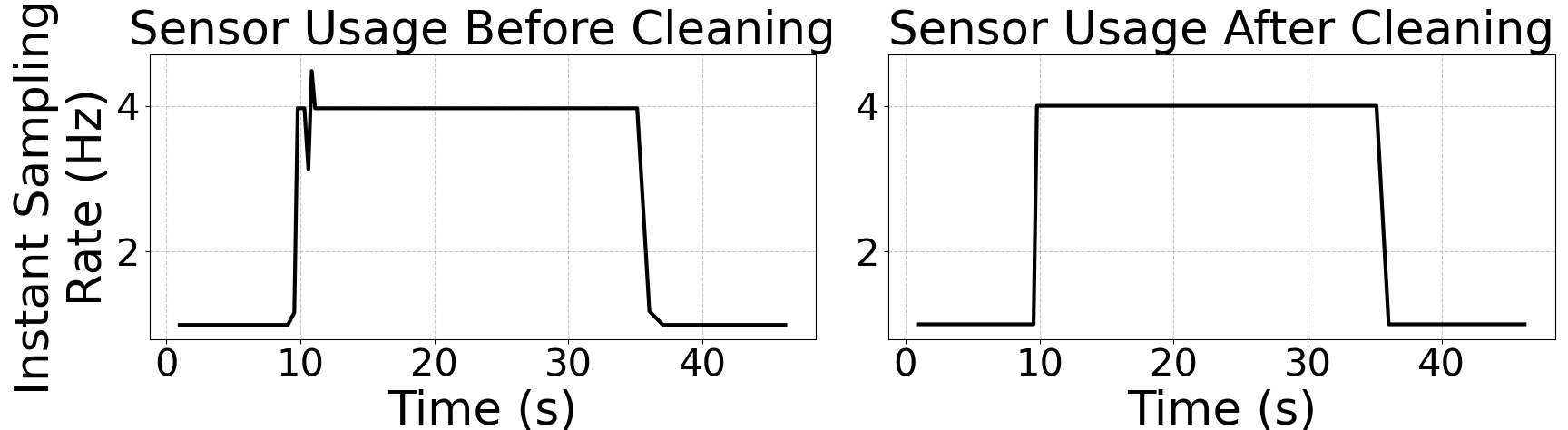}
  \vspace{-.15in}
  \caption{Example of the effect of outlier cleaning in $f_{inst}$.}
  \label{fig:outliter_cleaning}
  \vspace{-.2in}
\end{figure}

\begin{textblock*}{\textwidth}(2cm,13.4cm)
\noindent
\begin{minipage}{\textwidth}
  \centering
  \footnotesize
  \captionof{table}{Zero-permission Sensor Usage Pattern of Different Categories of Popular Google Play Apps}
  \label{table:percentages}
  \vspace{-0.15in}
  
  \begin{tabularx}{.9\textwidth}{c|c|c|c|c|c|X}
    \hline
    \textbf{Category (\# apps)} & \textbf{Accl.} & \textbf{Gyro.} & \textbf{Magn.} & \textbf{All} & \textbf{None} & \textbf{Examples of Apps Accessing Sensors} \\ 
    \hline\hline
    Games (38)            & 92\% & 63\% & 47\% & 47\% &  8\% & Roblox; Bus Out; Hole.io \\ \hline
    Art/Design (39)       & 67\% & 69\% & 33\% & 33\% & 28\% & \mline{Arvin® – AI Logo Maker; Creati AI Photo Generator; Sketchbook Lite} \\ \hline
    Finance (6)           & 67\% & 67\% & 67\% & 67\% & 33\% & Venmo; testerup – earn money; PayPal – Pay, Send, Save \\ \hline
    Family (35)           & 66\% &  6\% &  6\% &  6\% & 34\% & Bluey: Let's Play!; Floof – My Pet House; Disney Coloring World \\ \hline
    Personalization (47)  & 60\% & 62\% & 17\% & 17\% & 36\% & Launcher OS™; Easy Homescreen; Neon Love Theme \\ \hline
    Dating (37)           & 59\% & 57\% & 41\% & 41\% & 41\% & \mline{Dating \& Chat Online; Dating \& Chat – iHappy; Dating \& chat – Likerro} \\ \hline
    Photography (44)      & 57\% & 52\% & 30\% & 30\% & 43\% & Meitu; Skylight; Meta View \\ \hline
    Books/Reference (40)  & 55\% & 52\% & 42\% & 42\% & 45\% & WebNovel; NovelFlow; Holy Bible Light \\ \hline
    News/Magazines (48)   & 48\% & 48\% & 42\% & 40\% & 50\% & Newsmax; Quick News; News Today \\ \hline
    Education (35)        & 46\% & 34\% & 26\% & 23\% & 51\% & ClassDojo; Dino Coloring Game; Imprint: Learn Visually \\ \hline
    Entertainment (38)    & 45\% & 37\% & 29\% & 29\% & 55\% & Xbox; STARZ; Reel Rush \\ \hline
    Health/Fitness (39)   & 44\% & 44\% & 33\% & 33\% & 54\% & \mline{JustFit-Lazy Workout; Pilates Workout; Pedometer-Step Counter} \\ \hline
    Music/Audio (36)      & 42\% & 39\% & 19\% & 19\% & 56\% & Radio FM; Ringtones for Android™; Pocket FM: Audio Series \\ \hline
    Weather (44)          & 41\% & 39\% & 27\% & 27\% & 57\% & \mline{Weather\&Radar; Know Weather: Live Radar; Weather Forecasts\&Radar} \\ \hline
    Travel/Local (41)     & 39\% & 41\% & 46\% & 39\% & 54\% & earnify; Fly Delta; Disneyland® \\ \hline
    Beauty (47)           & 38\% & 38\% & 17\% & 15\% & 60\% & Barber Chop; GlossGenius; Picture Editor \\ \hline
    Social (39)           & 38\% & 33\% & 23\% & 23\% & 62\% & Facebook; Instagram; Letterboxd \\ \hline
    Parenting (47)        & 38\% & 28\% & 28\% & 28\% & 62\% & Alli360 by Kids360; Pregnancy Tracker: amma; Bark – Parental Controls \\ \hline
    Shopping (29)         & 38\% & 48\% & 45\% & 31\% & 48\% & Kroger; Lowe's; Circle K \\ \hline
    Comics (46)           & 37\% & 30\% & 17\% & 17\% & 63\% & K MANGA; Pocket Toons; Key Collector Comics \\ \hline
    Lifestyle (42)        & 36\% & 33\% & 29\% & 26\% & 62\% & AARP Now; Pinterest; Gold Town \\ \hline
    Video-Players (42)    & 36\% & 33\% & 10\% & 10\% & 64\% & Rumble; MX Player; Video Maker \\ \hline
    Maps/Navigation (47)  & 36\% & 34\% & 17\% & 17\% & 62\% & Phone Tracker; Roadie Driver; Bolt: Request a Ride \\ \hline
    Tools (46)            & 35\% & 35\% & 13\% & 13\% & 65\% & Manic; Wodfix Max; Neat Manager – AntiVirus \\ \hline
    House/Home (46)       & 35\% & 26\% & 17\% & 13\% & 65\% & Merkury Smart; Apartment List; PadSplit: Rooms for rent \\ \hline
    Sports (37)           & 32\% & 27\% & 22\% & 22\% & 68\% & NFL; GameChanger; NFL Network \\ \hline
    Events (44)           & 30\% & 25\% & 20\% & 18\% & 68\% & Timeleft; Posh – Social Experiences; Bridebook – Wedding Planner \\ \hline
    Food/Drink (40)       & 30\% & 22\% & 32\% & 18\% & 55\% & Wawa; Crumbl; Subway® \\ \hline
    Productivity (39)     & 26\% & 26\% &  5\% &  5\% & 74\% & AI Chatbot – Nova; PDF Reader – PDF Viewer; Email Lite – Smart Mail \\ \hline
    Auto/Vehicles (38)    & 24\% & 18\% & 18\% & 18\% & 76\% & PayByPhone; Fuel Forward; CARFAX Car Care App \\ \hline
    Medical (38)          & 21\% & 13\% & 11\% & 11\% & 79\% & Pathway; CSL Plasma; Sydney Health \\ \hline
    Libraries/Demo (47)   & 19\% & 15\% &  0\% &  0\% & 81\% & Cardboard; Samsung SmartTag; Addons for Melon \\ \hline
    Business (39)         & 18\% & 15\% & 10\% & 10\% & 82\% & Boat Browser; FedEx Mobile; Meta Business Suite \\ \hline
    Communication (41)    & 17\% & 15\% &  0\% &  0\% & 83\% & Messenger; Inbox Homescreen; Messenger – Texting App \\ \hline
    Android-Wear (37)     & 14\% & 14\% &  8\% &  8\% & 86\% & I am – Daily affirmations; CallApp: Caller ID\&Block; Fitbod \\ \hline
  \end{tabularx}

  \footnotesize
  *Different numbers of apps for each category (e.g., only six Finance apps) are due to DMCA, which prevented some APK/XAPK downloads.
\end{minipage}
\end{textblock*}

\end{document}